\documentclass[aps,prd,showpacs,showkeys, superscriptaddress, amssymb, amsmath, notitlepage, longbibliography, nofootinbib, natbib 10pt]{revtex4-1}
\allowdisplaybreaks
\usepackage{amsmath, amssymb, amsthm, gensymb, inputenc, multirow, breqn, dblfloatfix}
\usepackage{hyperref}
\makeatletter
\let\cat@comma@active\@empty
\makeatother
\DeclareMathOperator{\Tr}{Tr}

\newcommand{\be}{\begin{equation}}
\newcommand{\ee}{\end{equation}}
\newcommand{\ben}{\begin{eqnarray}}
\newcommand{\een}{\end{eqnarray}}
\newcommand{\nn}{\nonumber\\}

\begin{document}
\title{Electromagnetic and Axial Current Form Factors and Spectroscopy of Three-Flavor Holographic Baryons}
\author{Ori C. Druks}
\email{ori.druks@stonybrook.edu}
\affiliation{Department of Physics and Astronomy, Stony Brook University, Stony Brook, New York  11794-3800,  USA}
\author{Pak Hang Chris Lau}
\email{phcl2@mit.edu}
\affiliation{Center for Theoretical Physics, Massachusetts Institute of Technology, Cambridge, Massachusetts 02139, USA}
\author{Ismail Zahed}
\email{ismail.zahed@stonybrook.edu}
\affiliation{Department of Physics and Astronomy, Stony Brook University, Stony Brook, New York 11794-3800, USA}
\date{\today}
\begin{abstract}
We present an analysis of the three-flavor holographic model of QCD associated to a $D4/D8$ brane configuration, with symmetry breaking induced by a worldsheet instanton associated to a closed loop connecting $D4-D8-D6-\overline{D8}$.  We calculate the electromagnetic and axial couplings of all octet and decuplet baryons, as well as several negative parity excitations, with and without symmetry breaking effects, and demonstrate qualitative and quantitative agreement with many available experimental measurements, with marked improvement over the analogous two-flavor models.
\end{abstract}
\maketitle
\section{Introduction}
One of the hallmark of the holographic approach to QCD is the model proposed originally by Sakai and Sugimoto\cite{Sugim}, realizing a framework for the nonperturbative dynamics conceived by Witten\cite{Witt}, consisting of $N_f$ probe 
$D8 - \overline{D8}$ branes in a background of $N_c$ $D4$ branes. Baryons emerge as chiral solitons in the five dimensional Yang Mills-Chern-Simons theory resulting from the KK reduction on a circle. In the limit of large t'Hooft coupling, the instanton size is stabilized to a value on the order of the t'Hooft coupling to the inverse $1/2$ power, by the competing interaction of the outward-directed self-energy resulting from the Chern-Simons term and the inward-directed effects resulting from the curvature of the $D4$ color background. At correspondingly small size, the solution may be approximated by the flat space BPST instanton. The holonomy in the holographic direction yields a Skyrmion, following the Atiyah-Manton construction\cite{Atiyah}. 
\\ \indent By contrast to the original model of Skyrme\cite{Skyrme}, which neglects all mesons besides the massless pion, here an infinite tower of massive vector and axial-vector mesons is incorporated in a single $5d$ gauge field. The construction of the resulting baryonic currents provides a theoretical realization\cite{Zahed} of the empirically observed decomposition of the photon-hadron interaction into vector meson exchange (vector meson dominance).
\\ \indent The extension of the original models from $N_f=2$\cite{Sug} to $N_f = 3$ has been clouded for some time due to difficulty constructing an appropriate Chern-Simons term that both satisfies the requisite WZW constraint, required to correctly produce the baryon spectrum, as well as the chiral anomaly. The first problem was solved by Hata and Murata \cite{Hata}, at the expense of the second. Only recently, \cite{Chern}, a solution has been presented that satisfactorily solves both conditions.

The organisation  of the paper is as follows: in section II we detail the three flavor instanton construction and its quantization in bulk,
with particular emphasis on the role of the new Chern-Simon term. In section III we discuss the geometrical set up for the breaking of
chiral symmetry. In section IV, we construct the fully quantized vector and axial-vector currents for the three flavor baryons and their
respective form factors. In section V, we give detailed results for the electric and magnetic bulk baryon parameters without and with 
symmetry breaking effects.  In section VI and VII we discuss the charge radii and masses predictions for the octet and decuplet in
this holographic set up with comparison to existing models and lattice results.  In section VIII we analyze the axial charges of the excited
octet states, as well as their magnetic moments. Our conclusions are in section IX.

\section{Quantization}
Subtleties of the precise definition of the Chern-Simons term aside, the quantization of the flat space soliton, which appropriately approximates the curved space solution in the large t'Hooft coupling limit, is straightforward. We now recapitulate the salient details, clearly presented in \cite{Hata}. As noted, the fixed-time solution (with unit instanton number) is the Belavin-Polyakov-Schwarz-Tyupkin instanton. For $N_f=3$, we use the standard embedding

\be
A^{cl}_N (x) = -i f(\xi) g(x) \partial_N g(x)^{-1}\qquad{\rm where}\qquad  f(\xi) = \frac{\xi}{\xi^2 + \rho^2}, 
\ee
with $\xi = \sqrt{(x^N-X^N)^2}$ and

\begin{eqnarray} \begin{array}{cc}
g = \left(\begin{array}{cc} g^{SU(2)}(x) & 0 \\ 0 & 1 \end{array} \right), & g^{SU(2)} = \frac{1}{\xi} ((z-Z)I + i (x-X)^j \tau_j),
\end{array} \end{eqnarray}
$\tau_i$, $i=1,2,3$ are the Pauli matrices, normalized as  $\Tr(\tau_i \tau_j) = 2 \delta_{ij}$.

Following the moduli space approximation, the dynamics of the system, assumed to be slowly rotating in flavor space, is given by quantum mechanics on the instanton moduli space,  with coordinates given by the instanton position $X^N = (X, Z)$, size $\rho$, and $SU(3)$ orientation $V$, all assumed to be time-dependent, yet sufficiently slowly varying such that the dependence of the gauge field on the parameters is still given as in the static expression. The resulting Lagrangian has $\rho$-dependent moments of inertia. The principal difference of the $N_f=3$ case from $N_f=2$ is the introduction of an additional moment of inertia for the strange directions, with ratio $1/2$ to the moments corresponding to the original $SU(2)$ directions, as required to satisfy the Gauss law constraint.
Specifically, the Hamiltonian $H$ is given, up to corrections of order $1/\lambda \equiv 1/g^2 N_c$, by

\begin{eqnarray}
&&H = M_0 + H_Z + H_{\rho}\nn
&&H_Z  = \frac{1}{2m_Z} P^2_Z, \nn
&&H_{\rho} = -\frac{1}{2m_{\rho}} P^2_{\rho} + \frac{1}{2} m_{\rho} \omega_{\rho}^2 \rho^2 + \frac{K}{m_{\rho} \rho^2} 
+ \frac{1}{2 \mathcal{I}_1 (\rho)} \sum_{a=1}^{3} (J_a)^2 + \frac{1}{2 \mathcal{I}_2 (\rho)} \sum_{a=4}^{7} (J_a)^2 
\end{eqnarray}
where 

\begin{eqnarray}
&&P_Z^2 = - \frac{1}{2 m_z} \frac{\partial^2}{\partial Z^2},\nn
&&P_{\rho}^2 = - \frac{1}{2 m_{\rho}} \frac{1}{\rho^{\eta}} \frac{\partial}{\partial \eta} (\rho^{\eta} \partial_{\rho}) 
\end{eqnarray}
The inertial parameters read

\begin{eqnarray}
& & m_Z =  \frac 12 m_{\rho} = 8 \pi^2 a N_c, \nn 
& & K = \frac{N_c m_{\rho}}{40 \pi^2 a} = \frac{2}{5} N_c^2, \nn
& & \omega_z = \sqrt{\frac{2}{3}}, \omega_{\rho} = \sqrt{\frac{1}{6}}, \nn
& & \mathcal{I}_1 (\rho) = \frac{1}{4} m_{\rho} \rho^2, \ \mathcal{I}_2(\rho) = \frac{1}{2} \mathcal{I}_1 (\rho),
\end{eqnarray}
and the component of the angular momentum is defined as 

\begin{eqnarray}
J_a = - 4 \pi^2 i \kappa \Tr \left(  T_a V^{-1} \dot{V} \right),  \ a=1, \cdots, 7
\end{eqnarray}
 The effect of the Chern-Simons term is to impose the first-class constraint 
 \begin{eqnarray}
J_8 = \frac{N_c}{2 \sqrt{3}}.
\end{eqnarray}
As noted in \cite{Chern}, this is accomplished by defining the Chern-Simons action as

\begin{dmath}
S_{CS}  =  \frac{N_c}{24 \pi^2} \int_{M_5} \Tr \left( A F^2 - \frac{1}{2} A^3 F + \frac{1}{10} A^5 \right) + 2 \pi \alpha' \mu_8 \int_{D8/\overline{D8}} C_7 \wedge \Tr(F) + \frac{1}{10} \int_{N_5} \Tr ((\tilde{h}^{-1} d \tilde{h})^{5}) + \int_{\partial N_5} \alpha_4 (dh^{-1}h,A_{-}),
\end{dmath}
where the expression for $\alpha$ is given in \cite{Chern}.
The effect of the first class constraint that this generates, as noted, is important  in the construction of the resulting spectrum of eigenstates, as we shall now review.

The procedure of diagonalization is facilitated by examining the operator algebra satisfied by $J_a$, as well as $I_a \equiv - 4 \pi^2 \kappa \Tr(T^{a} V \dot{V}^{-1})$. As a consequence of the canonical commutation relations, the operators $J_a$ satisfy $[J_a, J_b] = i f_{abc} J_c$, and similarly $I_a$ satisfy $[I_a, I_b] = i f_{abc} I_c$, where $f_{abc}$ are the antisymmetric structure constants of $SU(3)$. By the completeness relation for $SU(3)$ generators,

\be
[J_a, V]=VT_a\qquad\qquad [I_a,V]=-T_a\nn
\ee
Since $I = VJV^{-1}$, and hence $\sum_{a=1}^8 I_a^2 = \sum_{a=1}^8 J_a^2$, the generators $I_a$ and $J_a$ arise from identical $SU(3)$ representations.
Explicit examination of the Noether currents reveals that $J_a$ is in fact minus the spin, the quantity conserved by rotational invariance, and $I_a$ is the flavor. Thus in summary, the system is quantized by spin and isospin generators in the same $SU(3)$ representation, a characteristic feature of Skyrme models.
\\ \indent The significance of the constraint induced by the Chern Simons term can now be appreciated. This is widely-known and well-reviewed, but we recapitulate for completeness. Following the standard $(p,q)$ parametrization of $SU(3)$ irreps (linear combinations of monomials with $p$ indices of one type and $q$ of another, such that the contraction of both is zero, graphically corresponding to polygonal weight diagrams of which $p$ and $q$ are the side lengths), the value of $Y_R \equiv J_8$ at the top of a multiplet is equal to $(p+2q)/3$, and the maximum corresponding $SU(2)$ eigenvalue occurring for states with a given value of $Y_R$ is equal to

\be
m_{\max} \Bigg\{  \begin{array}{cc} \frac{p}{2} + \frac{t}{2} & t \leq q \\ \frac{p}{2} - \frac{t}{2} + q & t \geq q  \end{array}\nn
\ee
where $t \leq p + q$ is an integer parametrizing the difference between the $Y_R$ value of the given state and that of the top level. Substituting the constraint yields accordingly for $B=1$ that $m_{\max}$ is half-integer valued when $N_C$ is odd, and integer-valued when $N_C$ is even, reproducing the properties of the $SU(6)$ quark model.
\\ \indent 
The diagonalization of the Hamiltonian is now transparent. The baryon wave functions are simply appropriately normalized $SU(3)$ Wigner D-functions, with one index equal to the flavor and the other equal to minus the spin, 
\begin{eqnarray} \Psi_B(G) = \sqrt{\dim (R)} (-1)^{J_3 + \frac{1}{2}} D^{(R)}_{I_3, -J_3}(G).
\end{eqnarray}
The eigenenergies are dependent on all representation indices and are given by 
\begin{eqnarray}
E(R, n_{\rho}, n_{Z}) = M_0 + \sqrt{ \frac{(\eta - 1)^2}{24} + \frac{N_c^2}{35} + \frac{m_{\rho}\rho^2 }{6 \mathcal{I}_2} C_2 (R) + m_{\rho} \rho^2\left( \frac{1}{2 \mathcal{I}_1} - \frac{1}{2 \mathcal{I}_2} \right) j(j+1) } + \sqrt{\frac{2}{3}}(n_{\rho} + n_{Z} + 1),
\end{eqnarray}
where $C_2(R)$ is the quadratic Casimir. 
 Before proceeding, we pause however to note an immediate observation, and concurrence with nature, which is that, notwithstanding subtleties to be noted regarding the extraction of numerical values of the mass from this analysis, the soliton mass formula above already produces an important result, which is that the lowest mass baryon multiplets are indeed the octet $\mathbf{8} = (1,1)$ with spin $1/2$, and the decuplet $\mathbf{10} = (3,0)$ with spin $3/2$.
Among the other irreps of particular importance for present purposes, are those other ones occurring in $\mathbf{8} \otimes \mathbf{8}$ and $\mathbf{8} \otimes \mathbf{10}$, for reasons to be noted. These are ${\mathbf{10}^*} = (0,3)$, $\mathbf{27} = (2,2)$, and ${\mathbf{35}} = (4,1)$.

\section{Symmetry Breaking}

The original model of Sakai and Sugimoto \cite{Sug} is one of zero mass pions, and thus zero (bare) mass quarks. Hashimoto  \cite{Hashimoto} subsequently addressed this problem by introducing additional $D6$ branes, parallel to the original color branes, but with non-zero separation, yielding massive $W$-bosons (with mass proportional to the separation distance, as is standard in the hypermultiplet), which give mass to the quarks of the original QCD via a vertex joining those quarks to a condensate of the new quarks arising from the open strings stretching between the $D6$ and $D8/\overline{D8}$. The corresponding interaction is mediated by a disk-shaped worldsheet connecting the $D4-D8-D6-\overline{D8}$, fixed at a single value of the time coordinate, (hence an instanton), with chiral symmetry breaking mediated by the development of a smooth throat connecting the $D8$ and $\overline{D8}$.

The effect of this deformation is to introduce an additional amplitude given by
\begin{eqnarray}
\delta S =  \frac{B}{2} \int d^4 x \mathcal{P} \Tr \left( M \left(\exp \left( -i \int_{-z_m}^{z_m} A_z \right) -I_3 \right) \right) + c.c. ,
\end{eqnarray}
where $z_m$ parametrizes the location of the $D6$ on the $D8/\overline{D8}$. 
For the BPST solution, the evaluation of this expression is aided by the observation that the $A_z$ component, 
$-i ({\xi^2}/{\xi^2+\rho^2}) g \partial_z g^{-1}$, is proportional to a fixed element of $\mathfrak{su}(3)$ for all values of $z$, hence the coupled set of differential equations that would otherwise have to be solved in order to evaluate the path ordered exponential 
(${\partial_t U} U^{-1} = AU)$), truncates to a single integral. Therefore we have

\begin{eqnarray}
\delta S = \frac{B}{2} \int d^4 x \Tr(M(U+U^{\dagger} - 2 I_3)), 
\end{eqnarray}
with

\be
U \equiv e^{i \pi h(r)}= \exp\Bigg\{i \pi \left( 1 - \frac{1}{\sqrt{1 + \rho^2/r^2}} \right)\Bigg\}
\ee
Upon substituting the $SU(3)$ Gell-Mann Oakes Renner relations and writing the mass matrix as 

\begin{eqnarray}
M 
= \frac{1}{3} (m_u + m_d + m_s) I_3 + \frac{1}{2} (m_u - m_d) T_3 + \frac{1}{2 \sqrt{3}} (m_u + m_d - 2 m_s) T_8,
\end{eqnarray}
we then obtain that
\begin{dmath}
\delta S =  \frac{B}{2} \frac{16 \pi}{3} \rho^3 \int dr r^2 (1-\cos h(r)) \Bigg(  m_{\pi \pm}^2 \cdot \Big(1 + 2 D_{88}^{(8)} \Big)
+   m_{K \pm}^2 \cdot \Big(1+ \sqrt{3} D_{38}^{(8)} - D_{88}^{(8)} \Big) 
 +  m_{K_0}^2 \cdot \Big(1 - \sqrt{3} D_{38}^{(8)} - D_{88}^{(8)} \Big) \Bigg). 
\end{dmath}
The  introduction of quark masses on the dynamics of the soliton reduces to a simple problem of the evaluation of a quantum mechanical perturbation on the moduli space of collective coordinates. By the elementary Rayleigh-Schr\"odinger procedure, the perturbed energies are obtained by appropriately weighted summations of the matrix elements of the perturbation, sandwiched between the unperturbed states, wave functions are obtained analogously, and operators are obtained by sandwiching the unperturbed expressions between the resulting wave functions. 

\begin{eqnarray}
|B, k \rangle = | B, k \rangle + \sum_{k \neq k'} |B, k' \rangle \frac{ \langle B, k' | H^{'} | B, k \rangle}{E_{k}^{(0)} - E_{k'}^{(0)}}. 
\end{eqnarray}
\\ \indent It is at this point that we utilize the observation noted earlier, about the tensor products of the irreps. Since $\mathbf{8} \otimes \mathbf{8} = \mathbf{1} \oplus \mathbf{8}_S \oplus \mathbf{8}_A \oplus \mathbf{10} \oplus \mathbf{10}^{*} \oplus \mathbf{27}$, the evaluation of the effects of the perturbation on the octet involves the calculation of matrix elements of the perturbation just defined, sandwiched between $\mathbf{8}$ and intermediate states in $\mathbf{10}^{*}$ and $\mathbf{27}$. (The unperturbed part is the matrix element between $8$ and $8$).  Likewise, since $\mathbf{8} \otimes \mathbf{10} = \mathbf{8} \oplus \mathbf{10} \oplus \mathbf{27} \oplus \mathbf{35}$, the evaluation of the effects of the perturbation on the decuplet is reduced to the evaluation of matrix elements of the perturbation, sandwiched between $\mathbf{10}$ and intermediate states among the set $\mathbf{27}, \mathbf{35}$.

\section{Currents, Form Factors}

A large number of static properties of the baryonic states can be obtained from a study of the electromagnetic and axial form factors. As observed \cite{Sterman}, perturbative QCD fails to properly reproduce experimental measurements thereof. A study of the properties resulting from the 2-flavor holographic framework is presented in \cite{Sugim}, and in the following, we extend this analysis to the more realistic 3-flavor framework, with appropriate modifications. Underpinning their analysis and ours, is the holographic prescription for computing currents, via the definition of external gauge fields (left and right), obtained by the respective classical solutions evaluated at the boundary values $z \rightarrow \pm \infty$. Accordingly, \cite{Sugim}, 

\begin{eqnarray} \mathbb{J}_{V \mu} = \mathbb{J}_{L \mu} + \mathbb{J}_{R \mu} = - (k(z) \mathcal{F}_{\mu z}^{cl})|_{z=- \infty}^{z = + \infty},
\end{eqnarray}
and 
\begin{eqnarray} \mathbb{J}_{A \mu} = \mathbb{J}_{L \mu} - \mathbb{J}_{R \mu} = - (k(z) \psi_0 (z) \mathcal{F}_{\mu z}^{cl})|_{z=-\infty}^{z = + \infty},
\end{eqnarray}
where $k(z) = 1 + z^2$ is a warp factor describing the curvature of the gravitational background, and $\psi_0(z) = \frac{2}{\pi} \arctan z$ is the zero mode in the KK-reduction of the gauge field, corresponding to the pion. 
 To obtain the gauge field at the boundary, we must gauge transform, since the BPST solution is singular there as originally written. Then we rewrite the resulting expressions for the boundary-valued field strengths in terms of Green's functions to the curved space wave equation, and expand these in a basis of the meson wave functions, as in \cite{Sugim}.

The analysis proceeds as for $SU(2)$, with the additional fact that the modification of the Gauss law constraint results in additional terms in the zero component of the gauge field for the $SU(3)$ generators $T_{a}$, $a=4, \cdots, 7$. \cite{Hata} %\footnote{As noted in \cite{Hata}, the constraint yields $D^{cl}_N D^{cl}_N U_{G} = 0$, or equivalently, $\frac{1}{\xi^3} \frac{d}{d \xi} (\xi^3 \frac{d}{d \xi} U^A) = k_A \frac{(1-f(\xi))^2}{\xi^2} U^A(\xi)$, where $k_A \Bigg\{  \begin{array}{cc} 0 & a = 8 \\
%8 & a= 1, 2, 3 \\ 3 & a = 4, 5,6,7 \end{array}$, which is solved by $U^8=1$, $U^{a} = f(\xi) = \frac{\xi^2}{\xi^2 + \rho^2}$ for $a=1,2,3$, and $U^a(\xi) = \sqrt{f(\xi)}$ for $a=4,\cdots 7$.}%
 The resulting terms in the expressions for the field strengths simply acquire an additional factor of $1/2$ compared to the $a=1, 2, 3$ terms %\footnote{This factor appears when taking the $\rho \ll \xi \ll 1$ limit, according to which  $\frac{\xi}{\sqrt{\xi^2 + \rho^2}} = \xi ( \frac{1}{\xi} - \frac{\rho^2}{2} \frac{1}{\xi^3} + \cdots) = 1 - \frac{\rho^2}{2} \frac{1}{\xi^2} + \cdots$, whereas $\frac{\rho^2}{\xi^2 + \rho^2} = \rho^2 \frac{1}{\xi^2 + \rho^2} = \rho^2 ( \frac{1}{\xi^2} - \frac{\rho^2}{\xi^4} + \cdots )$.}
, as is the case for the respective moments of inertia appearing in the collective coordinate Hamiltonian. For completeness, we note that 
\begin{eqnarray}
\mathbb{J}^{0}_{V/A} & = &
2 \pi^2 \kappa \left\{\partial_0 (\rho^2 a T^a a^{-1} ) \partial_a H_{V/A} - a T^a a^{-1} \rho^2 \dot{X}^i ((\partial_a \partial_i - \delta^{ai} \partial_j^2) H_{V/A} - \epsilon_{iaj} \partial_j G_{V/A}) \right. \nonumber \\ \nonumber
 & & \left. -  2  \rho^2 ((a_8 \dot{a}_a - a_a \dot{a}_8 + a_{a_1} a_{a_2} \epsilon_{a_1 a_2 a})T^a 
-  i (( \frac{1}{\sqrt{3}} (a_1 \dot{a}_8 + a_8 \dot{a}_1 + \cdots ))+ \cdots) \cdot G_{V/A}  \right. \nonumber
\\ & & \left. - \rho^2 ( (a_8 \dot{a}{_{\tilde{a}}} - a_{\tilde{a}} \dot{a}_8 )T^{\tilde{a}} + \cdots ) \cdot G_{V/A} \right \}.
\end{eqnarray}
The spatial currents are identical in form to $SU(2)$ \cite{Sug}, 
\begin{eqnarray}
\mathbb{J}^{i}_{V/A} = -2 \pi^2 \kappa \rho^2 a T^a a^{-1} ((\partial_i \partial_a - \delta^{ia} \partial_j^2) H^{V/A} - \epsilon^{ial}\partial_l G^{V/A} ),
\end{eqnarray}
where 

\begin{eqnarray}
&&G^V = (k(z) \partial_z G)|_{z=-\infty}^{z=+\infty}, \nn
&&G^A \equiv (\psi_0 (z) k(z) \partial_z G)|_{z=-\infty}^{z=+\infty}\nn
&&G(\vec{x}, z, \vec{X},Z) = \kappa \sum_{n=1}^{\infty} \psi_n (z) \psi_n(Z) Y_n (|\vec{x} - \vec{X}|)\,.
\end{eqnarray}
The expressions involving $H$ refer to the analogous Green's functions expanded in the basis $\phi_0 (z) = {1}/({\kappa \pi} {k(z)})$, $\phi_n(z) =\partial_z \psi_n/ {\sqrt{\lambda_n}}$.

The form factors are obtained by evaluating appropriate combinations of the Fourier-transformed currents in the basis of baryon wave functions. In particular, the elastic (Sachs) electric form factor $G_E$ is obtained in Breit frame (wherein the photon has zero energy) via the evaluation of the matrix elements of $J^0_V + J^8/{\sqrt{3}}$, and the magnetic $G_M$ from the spatial part $J_V^i +  J^8/{\sqrt{3}}$. This combination manifestly satisfies the Gell-Mann-Nishijima formula, $Q = I_3 + Y/2$. The axial form factors, meanwhile, are obtained from $J_A$. These carry a flavor index, and the appropriate choice is dictated by the decay under consideration, as will be clear.  
In terms of the Dirac and Pauli form factors $F_D$ and $F_P$, we have

\ben
&&G_E (\vec{k}^2) = F_D (\vec{k}^2) - \frac{k^2}{4 m_B^2} F_P (\vec{k}^2)\nn
&&G_M (\vec{k}^2) = F_D (\vec{k}^2) + F_P (\vec{k}^2)
\een
When taking the requisite Fourier transforms, we note as in \cite{Sug} that

\ben
&&\int d^3 x e^{-i \vec{k} \dot \vec{x}} Y_n (|\vec{x} - \vec{X}) = -e^{-i \vec{k} \dot \vec{X}} \frac{1}{\vec{k}^2 + \lambda_{n}}\nn
&&\int d^3 x e^{-i \vec{k} \dot \vec{x}} H_A (|\vec{x} - \vec{X}) = -e^{-i \vec{k} \dot \vec{X}} \frac{1}{\vec{k}^2} \sum_{n=1}^{\infty} \frac{g_{a^n} \partial_Z \psi_{2n} (Z)}{\vec{k}^2 + \lambda_{n}}
\een
\\ Accordingly, we find that for the positive-parity states,

\begin{eqnarray}
&&G_E (\vec{k}^2)= \kappa \langle \rho^2 \rangle Q \sum_{n \geq 1} \frac{g_{v^n} \psi_{2n-1}(Z)}{\vec{k^2} + \lambda_{2n-1}},\nn
&&G_M (\vec{k}^2)= - 8 \pi^2 \kappa \langle \rho^2 \rangle \Big(D_{33}^{(8)} + \frac{1}{\sqrt{3}} D_{83}^{(8)} \Big)\sum_{n \geq 1} \frac{g_{v^n} \psi_{2n-1}(Z)}{\vec{k^2} + \lambda_{2n-1}},
\end{eqnarray}
assuming unbroken $SU(3)$ symmetry. We note that for $\vec{k} = 0$, the expectation value of each respective expression, with respect to the baryon states, simply reduces to the quantity in parentheses, which is the electric charge in the first case, and the magnetic moment in the second. This follows from the fact that

\ben
\sum_{n=1}^{\infty} \frac{g_v^n}{\lambda_{2n-1}} \langle \psi_{2n-1}(Z) \rangle = 1\qquad{\rm with}\qquad
g_{v^n} = \lambda_{2n-1} \kappa \int dz h(z) \psi_{2n-1}(z)
\een
 and the completeness relation for the meson wave functions $\psi_n(z)$. The analogous procedure for the axial currents results in a factor of $\frac{2}{\pi} \langle \frac{1}{k(Z)} \rangle$. 
The ymmetry-breaking effects are obtained, as described earlier and to be done shortly, by a quantum mechanical perturbation theory calculation, with the perturbation given before.

The indices of the Wigner D-functions refer to the Cartesian basis. The mapping from this basis to Weyl-Cartan is such that $T_3 = (Y,I,I_3) = (0,1,0)$, $T_8 = (0,0,0)$. The combinations $T_4 + i T_5$ and $T_6 + i T_7$ respectively correspond to the isospin indices of $p$, $n$. Similarly, $T_4 - iT_5$, $T_6 - iT_7$ correspond to the isospin indices of $\Xi^-, \Xi^0$, and $T_1 - iT_2$, $T_1 + iT_2$ to the isospin indices of $\Sigma^{-}$, $\Sigma^{+}$. The identification of the spin states involves the minus sign noted earlier.
\\ \indent This identification of states is particularly important when considering the axial currents. These involve transitions between baryons, and the appropriate corresponding indices are dictated, as noted earlier, by the particular transition. For example, the transition $n \rightarrow p$ involves the current carrying an index $T_1 + iT_2$, corresponding to exchange of the meson $\pi^{-}$. By contrast, the transition $\Lambda \rightarrow p$ carries an index $T_4 + iT_5$, corresponding to exchange of $K^{-}$. 
The form factor for the first transition in the case of unbroken flavor symmetry, is given by
\begin{eqnarray}
(G_A)^{n \rightarrow p} = 8 \pi^2 \kappa \langle \rho^2  D_{1+i2, 3}^{(8)} \rangle \sum_{n \geq 1} 
\frac{g_{a^n} \partial_Z \psi_{2n}(Z)}{\vec{k}^2 + \lambda_{2n}},
\end{eqnarray}
and the corresponding result for the second transition is given by the same expression, but with the matrix element replaced with that of $D_{4+i5,3}$ between the requisite states. There are also flavor-neutral (i.e. diagonal) axial current elements, that do not change one baryon state to another. For each baryon, there is one such element corresponding to $3$ (associated to the current $\overline{u} \frac{T_3}{2} \gamma_{\mu} \gamma_5 u$) and one corresponding to $8$ (associated to the current $\overline{u} \frac{T_8}{2} \gamma_{\mu} \gamma_5 u$). These accordingly involve the matrix elements of $D_{3,3}^{(8)}$ and $D_{8,3}^{(8)}$.

 In addition, there are axial transitions that send one multiplet to another. Of particular note is the $n \rightarrow \Delta^{+}$ transition, which can be used to infer the value of the $\pi N \Delta$ coupling. These axial elements are calculated analogously to those just noted, with the only difference being that for the decuplet-to-octet transitions for example, the spin index in the $D$-function must be taken to be $1+i2$ rather than $3$.
The calculation of all such $D$-matrix elements is facilitated by the basic observation that they are equal to a product of $SU(3)$ Clebsch-Gordan coefficients, 

\begin{eqnarray}
\langle  D_{ab}^{(8)}  \rangle = \sum_{R} \left( \begin{array}{ccc} 8 & R_1 & R \\ a & \mu_1 & \mu \end{array} \right) \left( \begin{array}{ccc} 8 & R_1 & R \\ b & \nu_1 & \nu \end{array} \right).
\end{eqnarray}

\section{Results}

\subsection{Magnetic Moments: Unbroken}
 We now collect the results for the unbroken magnetic moments. We observe $U$-spin symmetry, as expected. Namely the unbroken moments are equal for states of equal electric charge. 
We have as follows:

\begin{eqnarray}
\mu_{p} & = &  \mu_{\Sigma^{+}}  = \frac{16 \pi^2}{15} \kappa \langle \rho^2 \rangle_{\mathbf{8}}  \nn
\mu_{\Sigma^{-}} & = & \mu_{\Xi^{-}} =  - \frac{4 \pi^2}{15} \kappa  \langle \rho^2 \rangle_{\mathbf{8}} \nn
\mu_n  & = &  \mu_{\Xi^{0}}  =  - \frac{4 \pi^2}{5} \kappa \langle \rho^2 \rangle_{\mathbf{8}} \nn
\mu_{\Sigma^0} &  = &  - \mu_{\Lambda^0}  =  \frac{4 \pi^2 \sqrt{5}}{15} \kappa \langle \rho^2 \rangle_{\mathbf{8}},
\end{eqnarray}
and 

\begin{eqnarray}
\mu_{\Delta^{++}} & = &  2 \pi^2 \kappa \langle \rho^2 \rangle_{\mathbf{10}}, \nn
\mu_{\Delta^{+}} & = & \mu_{\Sigma^{*+}}  =  \pi^2 \kappa \langle \rho^2 \rangle_{\mathbf{10}}, \nn
\mu_{\Delta^{0}} & = & \mu_{\Sigma^0} =\mu_{\Xi^{*0}} = 0, \nn
\mu_{\Delta^{-}} & = & \mu_{\Sigma^{*-}} = \mu_{\Xi^{*-}} = \mu_{\Omega^{-}} = - \pi^2 \kappa \langle \rho^2 \rangle_{\mathbf{10}},
\end{eqnarray}
where 

\begin{eqnarray} \langle \rho^2 \rangle_{\mathbf{8}} \equiv \langle \mathbf{8} | \rho^2 | \mathbf{8} \rangle =  \frac{1}{6}  \left(\sqrt{\frac{5}{6}}+ \frac{1}{2} \sqrt{\frac{689}{6}} \right) \rho_c^2, 
\end{eqnarray}
and 
\begin{eqnarray} \langle \rho^2 \rangle_{\mathbf{10}} \equiv \langle \mathbf{10} | \rho^2 | \mathbf{10} \rangle = 
\frac{1}{6} \left(\sqrt{\frac{5}{6}} + \frac{1}{2} \sqrt{\frac{929}{6}} \right) \rho_c^2. 
\end{eqnarray}
The numerical values are obtained in units of the Bohr nuclear magneton by multiplying with the factor $2M_N/ M_{KK}$, where $ M_{KK}$ is taken to be $949$ MeV. The results are summarized in Tables \ref{tab1} and \ref{tab2}.

A note is in order regarding the comparison of the decuplet quantities to the empirical data. Due to the fact that all decuplet particles other than $\Omega^{-}$ undergo strong interaction decays, and correspondingly have lifetimes on the order of $10^{-23}$ s, the measurement of their properties is an experimental challenge, and existing data sets, where available, have large errors, with the corresponding exception of $\Omega^-$. We note a good concordance of our prediction with this measured value in the last entry of Table \ref{tab2}. For this quantity and all others in the column, we also find agreement with the results of lattice simulations \cite{Leinweber}. 

\subsection{Axial Current Elements: Unbroken}

The matrix elements of the axial current display particularly improved agreement with empirical measurements, and particularly notable improvement over $SU(2)$. Here we present the expressions obtained for unbroken $SU(3)$, and later the results with symmetry breaking effects. The unbroken results display the expected $F$ and $D$ parametrization \cite{Gaillard} in the case of the octet, and the $C$, $H$ dependence in the case of the decuplet and the decuplet-to-octet transitions. 
The $\pi^{-}$ transitions for the octet are given by
\begin{eqnarray}
g_A^{np} & = & \frac{28 \sqrt{2}\pi}{15} \kappa \left\langle \frac{\rho^2}{k(Z)} \right\rangle_{\mathbf{8}} = F + D = 1.1478 \nn
g_A^{\Sigma- \Lambda^0} & = & \frac{4 \sqrt{3} \pi}{5} \kappa \left\langle \frac{\rho^2}{k(Z)} \right\rangle_{\mathbf{8}} = \frac{2D}{\sqrt{6}} = 0.60249 \nn
g_A^{\Sigma- \Sigma^0} &  = & \frac{4 \pi}{3}  \kappa \left\langle \frac{\rho^2}{k(Z)} \right\rangle_{\mathbf{8}} = \sqrt{2} F = 0.57974 \nn
g_A^{\Xi^{-} \Xi^{0}} & = & \frac{8 \sqrt{2} \pi}{15} \kappa \left\langle \frac{\rho^2}{k(Z)} \right\rangle_{\mathbf{8}} = D - F = 0.8116, 	
\end{eqnarray}
and the $K^{-}$ are given by 

\begin{eqnarray}
g_A^{\Lambda p} & = & \frac{16 \sqrt{3} \pi}{15} \kappa  \left\langle \frac{\rho^2}{k(Z)} \right\rangle_{\mathbf{8}} = \frac{3F + D}{\sqrt{6}} = 0.8095, \nn
g_A^{\Sigma^- n} & = &  D-F = 0.328, \nn
g_A^{\Xi^- \Lambda} & = & \frac{\sqrt{3} \pi}{30} \kappa \left\langle \frac{\rho^2}{k(Z)} \right\rangle_{\mathbf{8}} = \frac{3F-D}{\sqrt{6}} = 0.2008, \nn
g_A^{\Xi^- \Sigma^0} & = & \frac{7 \pi}{30} \kappa\left\langle \frac{\rho^2}{k(Z)} \right\rangle_{\mathbf{8}} = \frac{F+D}{\sqrt{2}} = 0.8116. 
\end{eqnarray}
These results are presented again in Table \ref{tab3}, along with the empirical data, with which there is an impressive consistency. Analogously, for the decuplet, we obtain

\begin{eqnarray}
(g_A^{\Delta^{++} \Delta^{++}})_3 & = & - 3 \pi \kappa \left\langle  \frac{\rho^2}{k(Z)}  \right\rangle_{\mathbf{10}} = H =-1.484 \nn
(g_A^{\Sigma^{*+} \Sigma^{*+}})_3 & = & - 2 \pi \kappa \left\langle  \frac{\rho^2}{k(Z)}  \right\rangle_{\mathbf{10}} = \frac{2}{3} H \nn
(g_A^{\Xi^{*0} \Xi^{*0}})_3 & = & - \pi \kappa \left\langle \frac{\rho^2}{k(Z)} \right\rangle_{\mathbf{10}} = \frac{1}{3} H \nn
g_A^{\Delta^{-} \Delta^{0}} & = & \sqrt{6} \pi \kappa \left\langle \frac{\rho^2}{k(Z)} \right\rangle_{\mathbf{10}} = - \frac{\sqrt{6}}{3} H \nn
g_A^{\Delta^{0} \Delta^{+}} & = & 2 \sqrt{2} \pi \kappa \left\langle \frac{\rho^2}{k(Z)}  \right\rangle_{\mathbf{10}} = - \frac{2 \sqrt{2}}{3} H \nn
g_A^{\Sigma^{*-} \Sigma^{*0}} & = & 2 \pi \kappa \left\langle \frac{\rho^2}{k(Z)} \right\rangle_{\mathbf{10}} = - \frac{2}{3} H\nn
(g_A^{\Delta^{++} \Delta^{++}})_8 & = & - (g_A^{\Xi^{*0} \Xi^{*0}})_8 =  - \frac{1}{2} (g_A^{\Omega^- \Omega^-})_8 = \frac{\sqrt{3}}{3} H  
\end{eqnarray}
for the strangeness-preserving transitions, and 
\\
\begin{eqnarray}
\begin{array}{ccc}
g_A^{\Xi^{*-} \Xi^{*0}} = \frac{1}{2} g_A^{\Delta^{0} \Delta^{+}}, & & 
g_A^{\Sigma^{*-} \Delta^{0}} = g_A^{\Xi^{*-} \Xi^{*0}}, \\
g_A^{\Sigma^{*0} \Delta^{+}} = g_A^{\Sigma^{*0} \Sigma^{*+}}, & & 
g_A^{\Sigma^{*+} \Delta^{++}} = g_A^{\Delta^{-} \Delta^{0}}, \\
g_A^{\Xi^{*0 \Sigma^{*+}}} = g_A^{\Delta^{0} \Delta^{+}}, & & 
g_A^{\Omega^{-} \Xi^{*0}} = g_A^{\Delta^{-} \Delta^{0}}
\end{array}
\end{eqnarray}
\\
for the strangeness-changing transitions. We also calculate transition elements for decuplet-to-octet processes, which as noted earlier can be obtained using similar arguments. Our results are as follows

\begin{eqnarray}
g_A^{\Delta^0 p}& = & 8 \pi \kappa \left\langle \frac{\rho^2}{k(Z)} \right\rangle_{\mathbf{10}} \langle \mathbf{8}, p | D_{1+i2,1-i2} | \mathbf{10}, \Delta^{0} \rangle = \sqrt{\frac{2}{3}} C = -0.9599 \nn
g_A^{\Delta^{-} n} & = & \frac{16 \sqrt{3}}{15} \pi \kappa b = \sqrt{2} C = -1.6626 \nn
g_A^{\Sigma^{*0} \Sigma^{+}} & = & \frac{8 \sqrt{6}}{15} \pi \kappa b = -\frac{C}{\sqrt{3}} = 0.6788 \nn
g_A^{\Sigma^{*-} \Lambda^{0}} & = & \frac{8 \sqrt{2}}{5} \pi \kappa b = \frac{C}{\sqrt{2}} = -1.1757 \nn
g_A^{\Xi^{*-} \Xi^0} & = & - g_A^{\Delta^0 p} \nn
g_A^{\Delta^+ n} & = & 8 \pi^2 \kappa \langle \rho^2 \rangle_{\mathbf{10}} \langle \mathbf{8} n |D_{1+i2, 1+i2} | \mathbf{10}, \Delta^{+} \rangle = g_A^{\Delta^0 p}.
\end{eqnarray}
Of particular note is the last quantity, which is in fact in good agreement with the value obtained from experimental measurement of the $\pi N \Delta$ coupling constant, $0.88$\cite{Semilepton}. 
\\
\\ \indent A comparison of our results for decuplet axial charges to empirical measurements, is challenged by the lack of data, just as for the magnetic moments. Accordingly, to substantiate our predictions, we perform a comparison with predictions obtained using other methods. We find notable agreement, in particular, with the results of a recent perturbative chiral quark model analysis (PCQM), \cite{Axial}, which also reproduces the measured octet axial charges, with better accuracy than either lattice methods, chiral perturbation theory, or relativistic chiral quark models. Our results for the octet charges are likewise in accord. %and share an agreement with experimental data that is superior to all the aforementioned models.

\subsection{Magnetic Moments with Symmetry Breaking}

The effects of symmetry breaking exhibit some properties of $V$-spin symmetry-- that is, symmetry along the right diagonals in the weight diagrams, as opposed to the left diagonals defining the $U$ spins. 
As noted earlier, the computation of the relevant quantities for the octet involves the mixing of $\mathbf{8}$ with $\mathbf{10^*}, \mathbf{27}$, and the computation of the quantities for the decuplet involves the mixing of $\mathbf{10}$ with $\mathbf{27}, \mathbf{35}$. Accordingly, we parametrize the results as follows. For the octet, 
\begin{eqnarray}
\mu_M & = & \mu_M^{(0)} + \delta \mu_M \nonumber \\ 
& = & \mu_M^{(0)} + \langle \mathbf{8} | \rho^2 | \mathbf{10^{*}}  \rangle \langle 8 |  \rho^3 | \mathbf{10^{*}}  \rangle \left( B^{10^{*}}_1  m_{K0, \overline{K0}} + B^{10^{*}}_2  m_{K \pm}^2 + B^{10^{*}}_3  m_{\pi \pm}^2 \right) \nonumber \\ 
& & + \langle \mathbf{8} | \rho^2 | \mathbf{27}  \rangle \langle \mathbf{8} | \rho^3 | \mathbf{27} \rangle \left( B^{27}_1  m_{K0, \overline{K0}} + B^{27}_2  m_{K \pm}^2 + B^{27}_3  m_{\pi \pm}^3 \right),
\end{eqnarray}
and similarly for the decuplet,
\begin{eqnarray}
\mu_M & = & \mu_M^{(0)} + \delta \mu_M \nonumber \\ 
& = & \mu_M^{(0)} + \langle \mathbf{10} | \rho^2 | \mathbf{27} \rangle \langle 10 | \rho^3 | \mathbf{27}  \rangle \left( B^{27}_1  m_{K0, \overline{K0}} + B^{27}_2  m_{K \pm}^2 + B^{27}_3  m_{\pi \pm}^2 \right) \nonumber \\ 
& & + \langle \mathbf{10} | \rho^2 | \mathbf{35} \rangle \langle \mathbf{10} \rho^3 | \mathbf{35} \rangle \left( B^{35}_1  m_{K0, \overline{K0}} + B^{35}_2  m_{K \pm}^2 + B^{35}_3  m_{\pi \pm}^3 \right).
\end{eqnarray}
In displaying the results, we list quantities for $V$-spin multiplets in successive order when possible, to exhibit the approximate symmetry noted.  The results for the decuplet are

\begin{eqnarray}
B_{27}^{\Delta^{++}} & = & 0.190515 \left( - \sqrt{3}, \sqrt{3} + \frac{\sqrt{30}}{8}, 3- \frac{\sqrt{30}}{8} \right) \nn
B_{27}^{\Sigma^{+}} & = & 0.186473 \left( - \sqrt{3} - \frac{1}{8} , \sqrt{3} + \frac{5}{8}, \frac{5}{2} \right) \nn
B_{27}^{\Xi^{*0}} & = & 0.189511 \left( - \sqrt{3} - \frac{\sqrt{6}}{12}, \sqrt{3} + \frac{5 \sqrt{6}}{24}, 3 - \frac{\sqrt{6}}{8} \right) \nn
B_{27}^{\Omega^{-}} & = & 0.06844 \left( - \sqrt{3} + \frac{17}{8}, \sqrt{3} - \frac{1}{8}, 1 \right) \nn
B_{35}^{\Delta^{++}} & = & 0.070336 \left( - \sqrt{3} + \frac{\sqrt{14}}{14}, \sqrt{3} + \frac{\sqrt{14}}{56}, 3 - \frac{5 \sqrt{14}}{56} \right) \nn
B_{35}^{\Sigma^{+}} & = & 0.070764 \left( - \sqrt{3} + \frac{3 \sqrt{35}}{56}, \sqrt{3} + \frac{\sqrt{35}}{56}, 3 - \frac{\sqrt{35}}{14} \right) \nn
B_{35}^{\Xi^{*0}} & = & 0.070608 \left( - \sqrt{3} + \frac{\sqrt{70}}{28}, \sqrt{3} - \frac{\sqrt{70}}{56}, 3 - \frac{\sqrt{70}}{56} \right) \nn
B_{35}^{\Omega^{-}} & = & 0.071985 \left( - \sqrt{3} + \frac{\sqrt{35}}{56} (2 + \sqrt{6}), \sqrt{3} + \frac{\sqrt{35}}{56} (2 - \sqrt{6}), 3 - \frac{\sqrt{35}}{14} \right) \nn
B_{27}^{\Delta^{+}} & = & 0.31194 \left( - \sqrt{3} + \frac{\sqrt{30}}{24}, \sqrt{3} + \frac{\sqrt{30}}{12}, 3 - \frac{\sqrt{30}}{8} \right) \nn
B_{27}^{\Sigma^{*0}} & = & 0.056899 \left( \frac{1}{4}, \frac{1}{4}, \frac{5}{2} \right) \nn
B_{27}^{\Xi^{*-}} & = & 0.056899 \left( - \sqrt{3} + \frac{5 \sqrt{6}}{24}, \sqrt{3} - \frac{\sqrt{6}}{12}, 3 - \frac{\sqrt{6}}{8} \right) \nn
B_{35}^{\Delta^{+}} &  = & 0.37348 \left( - \sqrt{3} + \frac{3 \sqrt{14}}{56}, \sqrt{3} + \frac{\sqrt{14}}{28}, 3 - \frac{5 \sqrt{14}}{56} \right) \nn
B_{35}^{\Sigma^{*0}} & = & 0.26723 \left( \frac{1}{4} \sqrt{\frac{5}{7}}, \frac{1}{4} \sqrt{\frac{5}{7}}, 3 - \frac{1}{2} \sqrt{\frac{5}{7}} \right) \nn
B_{35}^{\Xi^{*-}} & = & 0.026723 \left( - \sqrt{3} + \frac{\sqrt{70}}{56}, \sqrt{3} + \frac{\sqrt{70}}{28}, 3 - \frac{3 \sqrt{70}}{56} \right) \nn
B_{27}^{\Delta^0} & = & 0.172991 \left( - \sqrt{3} + \frac{\sqrt{30}}{12}, \sqrt{3} + \frac{\sqrt{30}}{24}, 3 - \frac{\sqrt{30}}{8} \right) \nn
B_{27}^{\Sigma^{-}} & = & 0.157679 \left( - \sqrt{3} + \frac{3}{8}, \sqrt{3} - \frac{3}{8}, \frac{5}{2} \right) \nn
B_{35}^{\Delta^0} & = & 0.069183 \left( - \sqrt{3} + \frac{\sqrt{14}}{28}, \sqrt{3} + \frac{3 \sqrt{14}}{56}, 3 - \frac{5 \sqrt{14}}{56} \right) \nn
B_{35}^{\Sigma^{*-}} & = & 0.068941 \left( - \sqrt{3} + \frac{\sqrt{35}}{36}, \sqrt{3} + \frac{3 \sqrt{35}}{56}, 3- \frac{\sqrt{35}}{14} \right) \nn
B_{27}^{\Delta^{-}} & = & 0.164229 \left( - \sqrt{3} + \frac{\sqrt{30}}{8}, \sqrt{3}, 3 - \frac{\sqrt{30}}{8} \right) \nn
B_{35}^{\Delta^{-}} & = & 0.068606 \left( - \sqrt{3} + \frac{\sqrt{14}}{56}, \sqrt{3} + \frac{\sqrt{14}}{14}, 3 - \frac{5 \sqrt{14}}{56} \right).
\end{eqnarray}
while for  the octet they are
 
\begin{eqnarray}
B_{10^{*}}^{n} & = & 0.100703 \left( - \sqrt{3} +  \frac{\sqrt{5}}{5} , \sqrt{3}, 3 - \frac{\sqrt{5}}{5} \right) \nn
B_{10^{*}}^{\Sigma-} & = & B_{10}^{n} \nn
B_{27}^{n} & = & 0.72406 \left( - \sqrt{3} + \frac{\sqrt{6}}{15} ,  \sqrt{3} + \frac{2 \sqrt{6}}{15}, 3 - \frac{\sqrt{6}}{5} \right) \nn
B_{27}^{\Sigma -} & = & 0.72591 \left( - \sqrt{3} + \frac{1}{5}, \sqrt{3} + \frac{1}{5}, \frac{13}{5} \right) \nn
B_{10^{*}}^{p} & = & 0.11121 \left( - \sqrt{3}, \sqrt{3} + \frac{\sqrt{5}}{5}, 3 - \frac{\sqrt{5}}{5} \right) \nn
B_{10^{*}}^{\Xi^-} & = & 0.72236 \left( \frac{\sqrt{6}}{10} - \frac{16 \sqrt{3}}{15}, \frac{\sqrt{6}}{10} + \frac{16 \sqrt{3}}{15}, 3 - \frac{\sqrt{6}}{5} \right) \nn
B_{27}^{p} & = & 0.07323 \left( - \sqrt{3} + \frac{2 \sqrt{6}}{15}, \sqrt{3} + \frac{\sqrt{6}}{15}, 3 - \frac{\sqrt{6}}{5} \right) \nn
B_{27}^{\Xi^-} & = & 0 \nn
B_{10^{*}}^{\Sigma^+} & = & 0.11121 \left( - \sqrt{3}, \sqrt{3} + \frac{\sqrt{5}}{5}, 3 - \frac{\sqrt{5}}{5} \right) \nn
B_{27}^{\Sigma^+} & = & B_{27}^{\Sigma -} \nn
B_{10^{*}}^{\Xi^0} & = & 0.0731 \left( - \sqrt{3} + \frac{\sqrt{6}}{15}, \sqrt{3} + \frac{2 \sqrt{6}}{15}, 3 - \frac{\sqrt{6}}{5} \right) \nn
B_{27}^{\Xi^0} & = & 0 \nn
B_{10^{*}}^{\Sigma^0} & = & 0.11506 \left( - \sqrt{3} + \frac{\sqrt{5}}{10} (1 - \sqrt{3}), \sqrt{3} + \frac{\sqrt{5}}{10}(1+ \sqrt{3}), 3 - \frac{\sqrt{5}}{5} \right) \nn
B_{27}^{\Sigma^0} & = & 0.07333 \left( - \frac{4 \sqrt{3}}{5}, \frac{4 \sqrt{3}}{5}, 3 \right) \nn
B_{10^{*}}^{\Lambda^0} & = & 0.11121 \left( - \sqrt{3}, \sqrt{3}, 3 \right) \nn
B_{27}^{\Lambda^0} & = & 0.0731 \left( - \sqrt{3} + \frac{3}{10}, \sqrt{3} + \frac{3}{10}, \frac{7}{10} \right).
\end{eqnarray}

\subsection{Axial Transition Constants with Symmetry Breaking}

We use a similar notation for the perturbative contributions to the axial couplings as for the perturbative contributions to the magnetic moments above, only with $B_{R}^{i}$, replaced with $A_{R}^{i}$ ($R$ labels the representation, $i$ the coefficient of $M_{i}$ as above). Our results for the octet to octet transitions are as follows

\begin{eqnarray}
A_{10^{*}}^{\Sigma^- \Sigma^0} & = & 0.002653 \left( - \sqrt{3} + \frac{\sqrt{5}}{5}, \sqrt{3}, 3 - \frac{\sqrt{5}}{5} \right) \nn
& & - 0.000442 \left( - \sqrt{3} + \frac{\sqrt{5}}{10} (1 - \sqrt{3}), \sqrt{3} + \frac{\sqrt{5}}{10} (1+ \sqrt{3}), 3 - \frac{\sqrt{5}}{5} \right) \nn
A_{10^{*}}^{\Xi^{-} \Xi^{0}} & = & 0 \nn
A_{27}^{\Sigma^- \Sigma^0} & = & 0 \nn
A_{27}^{\Xi^{-} \Xi^{0}} & = & 0.000294 \left( - \sqrt{3} + \frac{\sqrt{6}}{6} - \frac{14 \sqrt{3}}{15}, \sqrt{3} + \frac{7 \sqrt{6}}{30} + \frac{14 \sqrt{3}}{15}, 6 - \frac{2 \sqrt{6}}{5} \right) \nn
A_{10^{*}}^{\Sigma^{-} \Lambda^{0}} & = & 0.005812 \left( - \sqrt{3} + \frac{\sqrt{5}}{5}, \sqrt{3} + \frac{\sqrt{5}}{10} - \frac{\sqrt{15}}{30}, 3 - \frac{\sqrt{5}}{5} \right) \nn
A_{27}^{\Sigma^{-} \Lambda^{0}} & = & 0.000882 \left( - \sqrt{3} + \frac{3}{10}, \sqrt{3} + \frac{3}{10}, \frac{12}{5} \right) + 0.000147 \left( - \sqrt{3} + \frac{1}{5}, \sqrt{3} + \frac{1}{5}, \frac{13}{5} \right) \nn
A_{10^{*}}^{\Sigma^{-} n} & = & 0.003752 \left( - \sqrt{3} + \frac{\sqrt{5}}{5}, \sqrt{3}, 3 - \frac{\sqrt{3}}{5} \right) \nn
A_{27}^{\Sigma^{-} n} & = & 0.00072 \left( - \sqrt{3} + \frac{1}{5}, \sqrt{3} + \frac{1}{5}, \frac{13}{5} \right) 
- 0.000294 \left( - \sqrt{3} + \frac{\sqrt{6}}{15}, \sqrt{3}+ \frac{2 \sqrt{6}}{15}, 3 - \frac{\sqrt{6}}{5} \right). 
\end{eqnarray}
For the decuplet to decuplet transitions, the results are

\begin{eqnarray}
A_{27}^{\Delta^{-} \Delta^{0}} & = & 0.005417 \left( - 2 \sqrt{3} + \frac{5 \sqrt{30}}{24}, 2 \sqrt{3} + \frac{\sqrt{30}}{24}, 6 - \frac{\sqrt{30}}{4} \right) \nn
A_{35}^{\Delta^{-} \Delta^{0}} & = & 0.000356 \left( -2 \sqrt{3} + \frac{3 \sqrt{14}}{56}, 2 \sqrt{3} + \frac{\sqrt{14}}{8}, 6 - \frac{5 \sqrt{14}}{28} \right) \nn
A_{27}^{\Delta^{0} \Delta^{+}} & = & 0.006255\left( - 2 \sqrt{3} + \frac{\sqrt{30}}{8}, 2 \sqrt{3} + \frac{\sqrt{30}}{8}, 6 - \frac{\sqrt{30}}{4} \right) \nn
A_{35}^{\Delta^{0} \Delta^{+}} & = & 0.000411 \left( -2 \sqrt{3} + \frac{5 \sqrt{14}}{56}, 2 \sqrt{3} + \frac{5 \sqrt{14}}{56}, 6 - \frac{5 \sqrt{14}}{28} \right) \nn
A_{27}^{\Sigma^{*-} \Sigma^{*0}} & = & 0.007267 \left( -2 \sqrt{3} + \frac{7}{8}, 2 \sqrt{3} + \frac{1}{8}, 5 \right) \nn
A_{35}^{\Sigma^{*-} \Sigma^{*0}} & = & 0.00046 \left( -2 \sqrt{3} + \frac{3 \sqrt{35}}{56}, 2 \sqrt{3} + \frac{5 \sqrt{35}}{56}, 6 - \frac{\sqrt{35}}{7} \right) \nn
A_{27}^{\Sigma^{*-} \Delta^{0}} & = & -0.001713 \left( - \sqrt{3} + \frac{5}{8}, \sqrt{3} - \frac{1}{8}, \frac{5}{2} \right) \nn
& & - 0.003127 \left( - \sqrt{3} + \frac{\sqrt{30}}{12}, \sqrt{3} + \frac{\sqrt{30}}{24}, 3 - \frac{\sqrt{30}}{8} \right) \nn
A_{35}^{\Sigma^{*-} \Delta^{0}} & = & 0.000325 \left( - \sqrt{3} + \frac{\sqrt{35}}{56}, \sqrt{3} + \frac{3 \sqrt{35}}{56}, 3- \frac{\sqrt{35}}{14} \right) \nn
& & - 0.001029 \left( - \sqrt{3} + \frac{\sqrt{14}}{28}, \sqrt{3} + \frac{3 \sqrt{14}}{56}, 3 - \frac{5 \sqrt{14}}{56} \right) \nn
A_{27}^{\Sigma^{*0} \Delta^{+}} & = & -0.002422 \left( - \sqrt{3} + \frac{1}{4}, \sqrt{3} + \frac{1}{4}, \frac{5}{2} \right) \nn
& & - 0.004423 \left( - \sqrt{3} + \frac{\sqrt{30}}{24}, \sqrt{3} + \frac{\sqrt{30}}{12}, 3 - \frac{\sqrt{35}}{14} \right) \nn
A_{35}^{\Sigma^{*0} \Delta^{+}} & = & 0.00046 \left( - \sqrt{3} + \frac{\sqrt{35}}{28}, \sqrt{3} - \frac{\sqrt{35}}{28}, 3 - \frac{\sqrt{35}}{14} \right) \nn
& & - 0.001455 \left( - \sqrt{3} + \frac{3 \sqrt{14}}{56}, \sqrt{3} - \frac{\sqrt{14}}{28}, 3 - \frac{5 \sqrt{14}}{56} \right) \nn
A_{27}^{\Sigma^{*+ \Delta^{++}}} & = & 0.002967 \left( - \sqrt{3} - \frac{1}{8}, \sqrt{3} + \frac{5}{8}, \frac{5}{2} \right) \nn
& & - 0.005417 \left( - \sqrt{3}, \sqrt{3} + \frac{\sqrt{30}}{8}, 3 - \frac{\sqrt{30}}{8} \right) \nn
A_{35}^{\Sigma^{*+ \Delta^{++}}} & = & 0.000563 \left( - \sqrt{3} + \frac{3 \sqrt{35}}{56}, \sqrt{3} + \frac{\sqrt{35}}{56}, 3 - \frac{\sqrt{35}}{14} \right) \nn
& & - 0.001782 \left( - \sqrt{3} + \frac{\sqrt{14}}{14}, \sqrt{3} - \frac{\sqrt{14}}{56}, 3 - \frac{5 \sqrt{14}}{56} \right) \nn
A_{27}^{\Xi^{*-} \Sigma^{*0}} & = & -0.004845 \left( - \sqrt{3} + \frac{1}{4}, \sqrt{3} + \frac{1}{4}, \frac{5}{2} \right) \nn
& & - 0.001978 \left( - \sqrt{3} + \frac{5 \sqrt{6}}{24}, \sqrt{3} - \frac{\sqrt{6}}{12}, 3- \frac{\sqrt{6}}{8} \right) \nn
A_{35}^{\Xi^{*-} \Sigma^{*0}} & = & 0.000651 \left( - \sqrt{3} + \frac{\sqrt{70}}{56}, \sqrt{3} + \frac{\sqrt{70}}{28}, 3 - \frac{3 \sqrt{70}}{56} \right) \nn
& & - 0.00092 \left( - \sqrt{3} + \frac{\sqrt{35}}{28}, \sqrt{3} + \frac{\sqrt{35}}{28}, 3 - \frac{\sqrt{35}}{14} \right) \nn
A_{27}^{\Xi^{*0} \Sigma^{*+}} & = & -0.011648 \left( - \sqrt{3} + \frac{5}{8}, \sqrt{3} - \frac{1}{8}, \frac{5}{2} \right) \nn
& & - 0.002797 \left( - \sqrt{3} - \frac{\sqrt{6}}{12}, \sqrt{3} + \frac{5 \sqrt{6}}{24}, 3 - \frac{\sqrt{6}}{8} \right) \nn
A_{35}^{\Xi^{*0} \Sigma^{*+}} & = & 0.001807 \left( - \sqrt{3} + \frac{\sqrt{70}}{28}, \sqrt{3} - \frac{\sqrt{70}}{56}, 3 - \frac{3 \sqrt{70}}{56} \right) \nn
&  & - 0.001301 \left( - \sqrt{3} + \frac{\sqrt{35}}{56}, \sqrt{3} + \frac{3 \sqrt{35}}{56}, 3 - \frac{\sqrt{35}}{14} \right) \nn
A_{27}^{\Omega^{-} \Sigma^{*0}} & = & -0.007267 \left( - \sqrt{3} - \frac{\sqrt{6}}{12}, \sqrt{3} + \frac{5 \sqrt{6}}{12}, 3 - \frac{\sqrt{6}}{8} \right) \nn
A_{35}^{\Omega^{-} \Sigma^{*0}} & = & -0.000797 \left( - \sqrt{3} + \frac{\sqrt{70}}{28}, \sqrt{3} + \frac{\sqrt{70}}{56}, 3 - \frac{3 \sqrt{70}}{56} \right) \nn
& & + 0.001127 \left( - \sqrt{3} + \frac{\sqrt{35}}{56} ( 2 + \sqrt{6}), \sqrt{3} + \frac{\sqrt{35}}{56} ( 2- \sqrt{6}), 3 - \frac{\sqrt{35}}{14} \right).
\end{eqnarray}
Finally, for the decuplet-to-octet transitions, we obtain 

\begin{eqnarray}
A_{27, 3/2}^{\Delta^- n} & = & -0.00576 \left( - \frac{\sqrt{6}}{15} - \sqrt{3}, - \frac{2 \sqrt{6}}{15} + \sqrt{3}, 3 + \frac{\sqrt{6}}{10} \right) \nn
A_{27, 1/2}^{\Delta^- n} & = & -0.00654 \left( \frac{\sqrt{30}}{8} - \sqrt{3}, 0, 3 - \frac{\sqrt{30}}{8} \right) \nn
A_{27, 3/2}^{\Delta^0 p} & = & -0.00333 \left( \frac{\sqrt{6}}{15} - \sqrt{3}, \frac{2 \sqrt{6}}{15} + \sqrt{3}, 3 - \frac{\sqrt{6}}{5} \right) \nn
A_{27, 1/2}^{\Delta^0 p} & = & -0.00378 \left( \frac{\sqrt{30}}{12} - \sqrt{3}, \frac{\sqrt{30}}{24} + \sqrt{3}, 3 - \frac{\sqrt{30}}{8} \right)\nn
A_{27, 3/2}^{\Sigma^{0*} \Sigma^+} & = & 0 \nn
A_{27, 1/2}^{\Sigma^{0*} \Sigma^+} & = & 0.000865 \left( \frac{1}{5} - \sqrt{3}, \frac{1}{5} + \sqrt{3}, \frac{13}{5} \right) \nn
A_{27, 1/2}^{\Sigma^{-*} \Lambda^0} & = & -0.00666 \left( \frac{3}{10} - \sqrt{3}, \frac{3}{10} + \sqrt{3}, \frac{7}{10} \right) \nn
A_{27, 3/2}^{\Sigma^{-*} \Lambda^0} & = & -0.00507 \left( \frac{5}{8} - \sqrt{3}, - \frac{1}{8} + \sqrt{3}, \frac{5}{2} \right) \nn
A_{27, 1/2}^{\Xi^{*-} \Xi^0} & = & 0.001165 \left( \frac{7 \sqrt{6}}{15} - \sqrt{3}, \frac{2 \sqrt{6}}{15} + \sqrt{3}, 3 - \frac{\sqrt{6}}{5} \right) \nn
A_{27, 3/2}^{\Xi^{*-} \Xi^0} & = & 0.001689 \left( \frac{5 \sqrt{6}}{24} - \sqrt{3}, - \frac{\sqrt{6}}{12}, 3 - \frac{\sqrt{6}}{8} \right).
\end{eqnarray}

\section{Charge Radii}
The electric and magnetic charge radius of each baryon is defined, as usual, in terms of the first coefficient of the electric/magnetic form factor, expanded in powers of $\vec{k}^2$, 
\begin{eqnarray}
\langle r^2 \rangle_{E,M} = - \left. 6 \frac{d}{d \vec{k}^2} \ln G_{E,M}( \vec{k}^2 ) \right|_{\vec{k^2}=0}. 
\end{eqnarray}
Since the meson wave functions do not depend on flavor (as argued in \cite{Mass}, the effect of the worldsheet instanton perturbation on the vector meson mass is sub-dominant in the t'Hooft coupling), the expansion in question is the same as for $SU(2)$, and since the perturbation does not depend on the $Z$ coordinate, it does not modify the result. However, the analysis presented in \cite{Sug} for these quantities, does not include the decuplet baryons. It is noted therein that the results for the proton are in reasonable accord with empirical data, with some deviation, although the neutron is predicted to have vanishing radius on account of its vanishing charge, at variance with the small negative value that is measured. For the reasons just noted, this result does not change under the present considerations. We extend the analysis to the decuplet, however, and find an appealing agreement with the predictions of many different models, and in particular with field theoretical quark model (FTQM) calculations \cite{Sahoo} and with a $1/N_c$ analysis \cite{Lebed}. Following the framework just described, charge radii are expected to be the same for baryons with equal charge, and to differ for baryons of different charge, in proportion to the charge ratio. Accordingly, as we find that $\langle r^2 \rangle_{E,p} \approx (.784 \mbox{fm})^2$, we find that 
$\langle r^2 \rangle_{E, \Delta^{++}} \approx (1.109 \ \mbox{fm})^2$, compared to the FTQM prediction of $(1.086 \ \mbox{fm})^2$. The other predicted radii of the decuplet baryons are likewise in good accord with the predictions of the latter, which accordingly reproduce the expectation of electric charge proportionality, with small deviations therefrom in some cases. 
The results are summarized in Table \ref{tab6}.

\section{Mass Analysis}

The model that we use shares with general $SU(3)$ Skyrme models the effect of generating the mass relations of Gell-Mann- Okubo\cite{Gellmann} for the octet and decuplet baryons, as well as the relations of Coleman-Glashow\cite{Coleman}, and Guadagnini\cite{Guadagnini}. We start with an analysis up to first order in the perturbation, and treat the next-to -leading order correction thereafter. From the evaluation of the matrix elements $\langle B, \mathbf{8}| D_{38}^{(8)} | B, \mathbf{8} \rangle$ and $\langle B, \mathbf{8} | D_{88}^{(8)} | B, \mathbf{8} \rangle$, we find concurring with \cite{HashimIuzuka}, that

\begin{eqnarray}
M_{\Lambda} & = & M_{\mathbf{8}}+ \langle \Lambda, \mathbf{8} | H' |\Lambda, \mathbf{8} \rangle = M_8 + \langle \mathbf{8} | \rho^3 | \mathbf{8} \rangle c \left( \frac{9}{10}   m_{K0, \overline{K0}}^2 + \frac{9}{10}   m_{K \pm}^2 + \frac{6}{5}   m_{\pi \pm}^2 \right) \nn
M_{\Sigma_0} & = & M_{\mathbf{8}} + \langle \Sigma^0, \mathbf{8} | H' | \Sigma^0, \mathbf{8} \rangle = M_8 + \langle \mathbf{8} | \rho^3| \mathbf{8} \rangle c \left( \frac{11}{10}   m_{K0, \overline{K0}}^2 + \frac{11}{10}   m_{K \pm}^2 + \frac{4}{5}   m_{\pi \pm}^2 \right) \nn
M_{\Xi_0} & = & M_{\mathbf{8}} + \langle \Xi^0, \mathbf{8} | H' | \Xi^0, \mathbf{8} \rangle = M_8 + \langle \mathbf{8} | \rho^3 | \mathbf{8} \rangle c \left( \frac{4}{5}   m_{K0, \overline{K0}}^2 + \frac{8}{5}   m_{K \pm}^2 + \frac{3}{5}   m_{\pi \pm}^2 \right) \nn
M_{N} & = & M_{\mathbf{8}} + \langle N, \mathbf{8} | H' | N, \mathbf{8} \rangle = M_8 + \langle \mathbf{8} | \rho^3 | \mathbf{8} \rangle c  \left( \frac{4}{5}  m_{K_0}^2 + \frac{3}{5}   m_{K \pm}^2 + \frac{8}{5}   m_{\pi \pm}^2 \right),
\end{eqnarray}
where $M_{\mathbf{8}}$ denotes the soliton mass in the $\mathbf{8}$ representation, i.e. the corresponding energy eigenvalue of the collective coordinate Hamiltonian, 
\begin{eqnarray} 
M_{\mathbf{8}} = M_0 + \sqrt{\frac{(\eta-1)^2}{24} + \frac{1}{3}\left(\frac{N_c^2}{15} + 4 C_2(R) -2 j(j+1) \right)} + \frac{2}{3} (n_{\rho} + n_{z} + 1) = 8 \pi^2 \kappa +  \sqrt{\frac{137}{24}} + \sqrt{\frac{2}{3}}
\end{eqnarray}
(it units of $ M_{KK}$), and $c=1.104/3$.
Accordingly, eliminating $M_8$, we find that 
\begin{eqnarray}
3 M_{\Lambda} + M_{\Sigma^0} - 2(M_N + M_{\Xi^0}) = \frac{3}{5} c \langle \mathbf{8} | \rho^3 | \mathbf{8} \rangle  (  m_{K0, \overline{K0}}^2 -   m_{K \pm}^2),
\end{eqnarray}
which for the values $  m_{K0, \overline{K0}} = 498$ MeV, $ m_{K\pm} =494$ MeV, $ m_{\pi \pm} = 140$ MeV, is numerically equal to $6.269$ MeV, compared to an empirically measured value of $26$ MeV. The decuplet equal spacing rule of Gell-Mann Okubo, to first order, is satisfied exactly. To wit, 

\begin{eqnarray}
M_{\Delta^-} & = & M_{\mathbf{10}} + \langle \mathbf{10} | \rho^3 | \mathbf{10} \rangle c \left( \frac{5}{4}   m_{K0, \overline{K0}}^2 + \frac{1}{2}  m_{K \pm}^2 + \frac{5}{4}  m_{\pi \pm}^2 \right)\nn
M_{\Sigma^{*-}} & = & M_{\mathbf{10}} + \langle \mathbf{10} | \rho^3 | \mathbf{10} \rangle c \left( \frac{5}{4}   m_{K0, \overline{K0}}^2 + \frac{3}{4}  m_{K \pm}^2 + \frac{5}{4}  m_{\pi \pm}^2 \right) \nn
M_{\Xi^{* - }} & = & M_{\mathbf{10}} + \langle \mathbf{10} | \rho^3 | \mathbf{10} \rangle c \left( \frac{5}{4}   m_{K0, \overline{K0}}^2 +  m_{K \pm}^2 + \frac{3}{4}  m_{\pi+}^2 \right) \nn
M_{\Omega^-} & = & M_{\mathbf{10}} +  \langle \mathbf{10} | \rho^3 | \mathbf{10} \rangle c \left( \frac{5}{4}   m_{K0, \overline{K0}}^2 + \frac{5}{4}  m_{K \pm}^2 + \frac{1}{2}  m_{\pi \pm}^2 \right) 
\end{eqnarray}
hence 

\begin{eqnarray}
M_{\Delta^-} - M_{\Sigma^{*-}} = M_{\Sigma^{*-}} - M_{\Xi^{*-}} = M_{\Xi^{*-}} - M_{\Omega-} = \frac{1}{4} c \langle \mathbf{10} | \rho^3 | \mathbf{10} \rangle  \left(  m_{\pi \pm}^2 -  m_{K \pm}^2 \right),
\end{eqnarray}
where following the notation earlier, $M_{\mathbf{10}}$ denotes the mass of the soliton in the $\mathbf{10}$ representation.
The relation of Coleman-Glashow for the baryon octet, $M_P - M_N = (M_{\Sigma^+} - M_{\Sigma^-}) - (M_{\Xi^0} - M_{\Xi^-})$, is satisfied exactly as well to first order. Namely, 

\begin{eqnarray}
M_P = M_{\mathbf{8}} + \langle \mathbf{8} | \rho^3 | \mathbf{8} \rangle c \left( \frac{3}{5}   m_{K0, \overline{K0}}^2 + \frac{4}{5}  m_{K \pm}^2 + \frac{8}{5}  m_{\pi \pm}^2 \right) \nn
M_{\Sigma^+} = M_{\mathbf{8}} + \langle \mathbf{8} | \rho^3 | \mathbf{8} \rangle c \left( \frac{3}{5}   m_{K0, \overline{K0}}^2 + \frac{8}{5}  m_{K \pm}^2 + \frac{4}{5}  m_{\pi \pm}^2 \right) \nn
M_{\Sigma^-} = M_{\mathbf{8}} + \langle \mathbf{8} | \rho^3 | \mathbf{8} \rangle c \left( \frac{8}{5}   m_{K0, \overline{K0}}^2 + \frac{3}{5}  m_{K \pm}^2 + \frac{4}{5}  m_{\pi \pm}^2 \right) \nn
M_{\Xi^-} = M_{\mathbf{8}} + \langle \mathbf{8} | \rho^3 | \mathbf{8} \rangle c \left( \frac{8}{5}   m_{K0, \overline{K0}}^2 + \frac{4}{5}  m_{K \pm}^2 + \frac{3}{5}  m_{\pi \pm}^2 \right),
\end{eqnarray}
thus

\begin{eqnarray}
M_P - M_N = (M_{\Sigma^+} - M_{\Sigma^-}) - (M_{\Xi^0} - M_{\Xi^-}) = \frac{1}{5}  \langle \mathbf{8} | \rho^3 | \mathbf{8} \rangle c \left( m_{K \pm}^2 -   m_{K0, \overline{K0}}^2 \right).
\end{eqnarray}
Lastly, the relation of Guadagnini, $M_{\Xi^*} - M_{\Sigma^*} + M_N = \frac{1}{8} (11 M_{\Lambda} - 3 M_{\Sigma}) $, is satisfied with a small deviation. Namely, by the same analysis as above, the difference of the LHS from the RHS of the last equation is found to be 
\begin{eqnarray}
\frac{1}{40} c\langle \mathbf{8} | \rho^3 | \mathbf{8} \rangle  m_{K0, \overline{K0}}^2 + c \Bigg( \frac{9}{40} \langle \mathbf{8} | \rho^3 | \mathbf{8} \rangle  - \frac{1}{4}  \langle \mathbf{10} | \rho^3 | \mathbf{10} \rangle  \Bigg)  m_{K \pm}^2 + \frac{1}{4}c \Bigg(\langle \mathbf{10} | \rho^3 | \mathbf{10} \rangle  - \langle \mathbf{8} | \rho^3 | \mathbf{8} \rangle \Bigg)  m_{\pi \pm}^2, 
\end{eqnarray}
which numerically evaluates to $-30.7$ MeV.

 We now consider the effects of the next-to-leading order in the perturbation, % $\delta^{(2)} M_B = \sum_{b \neq b'} \frac{| \langle k',B | H' | k, B \rangle |^2}{E_k^{(0)} - E_{k'}^{(0)}}$.%
For compactness, %we evaluate $ \langle \mathbf{8} | \rho^3 | \mathbf{10^{*}} \rangle$ and $\langle \mathbf{8} | \rho^3 \mathbf{27} \rangle$,% 
we write the results only in terms of $ m_{K0, \overline{K0}}$, $ m_{K \pm}$, $ m_{\pi \pm}$.
We obtain as follows for the octet, suppressing a factor of $10^{-8}$

\begin{eqnarray}
\delta^{(2)} M_{\Sigma^{0}} & = & -1.1058 \left( - \frac{4 \sqrt{3}}{5}  m_{K0 \overline{K0}}^2+ \frac{4 \sqrt{3}}{5}  m_{K \pm} + 3  m_{\pi \pm}^2 \right)^2 \nn & & - 1.43641 \left( (-\sqrt{3} + \frac{\sqrt{5}}{10} (1 - \sqrt{3}))  m_{K0 \overline{K0}}^2 + \left( \sqrt{3} + \frac{\sqrt{5}}{10} (1 + \sqrt{3})\right)  m_{K \pm}^2 + \left( 3 - \frac{1}{\sqrt{5}} \right)  m_{\pi \pm}^2 \right)^2\nn
\delta^{(2)} M_{\Xi^{0}} & = & -1.1058 \left( \left( \frac{\sqrt{6}}{15} - \sqrt{3}\right)  m_{K0 \overline{K0}}^2 + \left( \frac{2 \sqrt{6}}{15} + \sqrt{3} \right)  m_{K \pm}^2 + \left( 3- \frac{\sqrt{6}}{5} \right) m_{\pi \pm}^2 \right)^2 \nn
\delta^{(2)} M_{\Lambda^{0}} & = & -1.1058 \left( \left( -\sqrt{3} + \frac{3}{10}\right)  m_{K0 \overline{K0}}^2 + \left( \sqrt{3} + \frac{3}{10}\right)  m_{K \pm}^2 + \frac{7}{10}  m_{\pi \pm}^2 \right)^2 \nn & & - 1.43461 \left( - \sqrt{3}  m_{K0 \overline{K0}}^2 + \sqrt{3}  m_{K \pm}^2 + 3  m_{\pi \pm}^2 \right)^2 \nn
\delta^{(2)} M_{N} & = & -1.1058 \left( \left( - \sqrt{3} + \frac{\sqrt{6}}{15} \right)  m_{K0 \overline{K0}}^2 + \left( \sqrt{3} + \frac{2 \sqrt{6}}{15}\right)  m_{K \pm}^2 + \left(3 - \frac{\sqrt{6}}{5} \right) m_{\pi \pm}^2\right) \nn
& & - 1.43641 \left( \left(\sqrt{3} - \frac{1}{\sqrt{5}}\right)  m_{K0 \overline{K0}}^2 +  3  m_{K \pm}^2  + \left(3 - \frac{1}{\sqrt{5}} \right)  m_{\pi \pm}^2 \right)^2.
\end{eqnarray}
Numerically, these expressions evaluate to $\delta^{(2)} M_{\Sigma^0} = -361.752$ MeV, $\delta^{(2)} M_{\Xi^0} = -292.093$ MeV, $\delta^{(2)} M_{\Lambda^0} = -302.549$ MeV, $\delta^{(2)} M_{N} = -340.983$ MeV, resulting in a net additional deviation from Gell-Mann-Okubo of $-3.249$ MeV, considerably smaller than the first-order correction.

 Likewise, for the decuplet, the second-order mass corrections are given by

\begin{eqnarray}
\delta^{(2)} M_{\Delta^{-}} & = & -2.906 \left( \left(-\sqrt{3} + \frac{\sqrt{30}}{8} \right) m_{K0 \overline{K0}}^2 + \sqrt{3} m_{K \pm}^2 + \left(3 - \frac{\sqrt{30}}{4}\right) m_{\pi \pm}^2 \right)^2\nn
& & - 1.2652 \left( \left(-\sqrt{3} + \frac{\sqrt{14}}{56}\right) m_{K0 \overline{K0}}^2 + \left( \sqrt{3} + \frac{\sqrt{14}}{14}\right)m_{K \pm}^2 + \left(3 - \frac{5 \sqrt{14}}{56}\right) m_{\pi \pm}^2 \right)^2 \nn
\delta^{(2)} M_{\Omega^{-}} & = & -2.9064 \left( \left(-\sqrt{3} + \frac{17}{8}\right) m_{K0 \overline{K0}}^2 + \left(\sqrt{3} - \frac{1}{8}\right)m_{K \pm}^2 + m_{\pi \pm}^2 \right)^2 \nn
& & -1.2652 \left( \left(-\sqrt{3} + \frac{\sqrt{35}}{28}(1-\sqrt{\frac{3}{2}})\right)m_{K0 \overline{K0}}^2
+ \left(\sqrt{3} + \frac{\sqrt{35}}{28} (1 + \sqrt{\frac{3}{2}})\right)m_{K \pm}^2 + \left(3 - \frac{\sqrt{35}}{2}\right) m_{\pi \pm}^2 \right)^2 \nn
\delta^{(2)} M_{\Xi^{*-}} & = & 1.2652 \left(\left( -\sqrt{3} + \frac{\sqrt{70}}{56} \right) m_{K0 \overline{K0}}^2 + \left(\sqrt{3} + \frac{\sqrt{70}}{28}\right) m_{K \pm}^2 + \left(3 - \frac{3 \sqrt{70}}{56}\right) m_{\pi \pm}^2 \right)^2 \nn
& & - 2.9064 \left( \left(-\sqrt{3} + \frac{5 \sqrt{6}}{24}\right) m_{K0 \overline{K0}}^2 + 
\left(\sqrt{3} - \frac{\sqrt{6}}{12}\right) m_{K \pm}^2 + \left(3 - \frac{\sqrt{6}}{8}\right) m_{\pi \pm}^2 \right)^2 \nn
\delta^{(2)} M_{\Sigma^{*-}} & = & -305.14 \left( \left(- \sqrt{3} + \frac{5}{8}\right) m_{K0 \overline{K0}}^2 + \left(\sqrt{3} - \frac{1}{8}\right) m_{K \pm}^2 + \frac{5}{2} m_{\pi \pm}^2 \right)^2\nn
& & - 111.42 \left( \left(- \sqrt{3} + \frac{\sqrt{35}}{56}\right) m_{K0 \overline{K0}}^2 + \left(\sqrt{3} + \frac{3 \sqrt{35}}{56}\right) m_{K \pm}^2 + \left(3 - \frac{\sqrt{35}}{28} \right) m_{\pi \pm}^2 \right)^2,
\end{eqnarray}
suppressing a factor of $10^{-3}$. Using these expressions, we accordingly obtain the Okubo relation \cite{Okubo}, $M_{\Omega} - M_{\Delta} = 3(M_{\Xi^{*}} - M_{\Sigma^{*}})$, with a small violation. Namely, with the meson masses quoted earlier, we find that the ratio of the LHS to the RHS in this relation is $1.0636$. 
\\ \indent The results of this analysis-- the generation of the aforementioned sum rules with violations in accord with empirical data-- confirm the consistency of the quantization. As noted in \cite{Hashimoto}, however, this analysis is not satisfactory for a prediction of numerical mass values, for which it is anticipated that the effects of flavor symmetry violation should not only include the leading-order disk worldsheet instanton considered above, but also worldsheets of higher instanton number. Higher-order string loops are also expected to yield higher order corrections in $1/N_c$.  Notwithstanding, it is noteworthy, as has been remarked by \cite{Baryon}, that the mass formula obtained at lowest order has the same $1/N_c$ dependence that is expected from a diagrammatic expansion of QCD, \cite{Luty}. Namely, the leading mass difference between ground state baryons of different spins is of order $1/N_c$ and is proportional to $J^2$.

\section{Excited States}

Among the virtues of the Sakai Sugimoto formalism is the simplicity with which it incorporates excited baryons of both even and odd parity. The baryon states considered heretofore in this paper are those corresponding to the zero value of the principal quantum number $n_{\rho}$ of the radial coordinate in the collective coordinate Hamiltonian, as well as the zero value of the quantum number $n_z$ corresponding to the z-coordinate. The wave function for the $z$ coordinate is simply that of a harmonic oscillator, and it accordingly has negative parity under $z \rightarrow - z$, for odd $n_z$. The effect of non-zero $n_{\rho}$, meanwhile, is to multiply the result for $n_{\rho}=0$, by the hypergeometric function

\be
{}_1 F_1 \left(-n_{\rho}, \beta_{(p,q)l}; 16 \pi^2 \kappa/\sqrt{6} \rho^2 \right)\qquad{\rm where}\qquad
\beta_{(p,q)l} = 1 + \sqrt{\frac{(\eta-1)^2}{4} + 2KÕ}\nn  
\ee
and $K'=\frac{N_c^2}{15} + 4C_2(R)-2j(j+1)$.
Accordingly, the static properties of the resulting states can be just as easily calculated as those for the ground state baryons considered earlier, simply with the suitable recalculation of the expectation values $\langle \rho^2 \rangle$ and $\langle \rho^3 \rangle$. The values of the axial charges for some of these states are tabulated in Table \ref{tab7}, and compared to predictions from a relativistic chiral quark model, \cite{Excited}. The states are identified with entries in the PDG essentially according to \cite{Baryon}. As in \cite{Sug}, the first positive parity excited state of the nucleon, with quantum numbers $(n_{\rho}, n_z) = (1, 0)$, is identified with the Roper excitation, $N(1440)$. As in \cite{Sug}, we find that the corresponding axial charge is larger than that of the neutron, however we find a smaller relative ratio ($1.29$ vs. $1.45$), in better agreement with the predictions of \cite{Excited}. Likewise as in \cite{Sug}, we find that first negative parity nucleon, with quantum numbers $(n_{\rho}, n_z)=(0,1)$, has smaller charge, and the ratio is improved in comparison to \cite{Excited}, yet the charge is larger. The next negative parity state, $(1,1)$, which we identify with $N^*(1655)$, has an axial charge that is larger, yet significantly smaller than that of the neutron, again in agreement with \cite{Excited}. Provided we identify the subsequent positive parity state $(0,2)$ as $N(1710)$, we again obtain a consistent result: the charge is smaller still than the preceding, but not by much.\
\\ \indent There is, however, a potential problem with the magnetic moments. One feature of the Sakai-Sugimoto formalism, observed in \cite{Sug}, although not remarked as a potential issue, is a degeneracy of the values thereof for negative parity baryons, with the ground state values. This is an intrinsic feature, for the following reason. As noted above, the only difference between the (lowest) negative parity excitations and the ground state wave functions is in the dependence on the $z$-variable. The lowest positive parity excitation also contains a change in the wave function $\rho$ coordinate, as do the higher excitations, however this is not true for the lowest negative parity state. The dependence on $z$ factors out of the expression for the magnetic moment, due to the form of the expression for the vector meson decay constants, $g_v^n  = \lambda_{2n-1} \kappa \int dz h(z) \psi_{2n-1}(z)$. To wit, recall that 
\begin{eqnarray}
G_M (\vec{k}^2) \propto \langle \rho^2 \rangle \sum_{n \geq 1} \frac{g_{v^n} \psi_{2n-1}(Z)}{\vec{k^2} + \lambda_{2n-1}}. 
\end{eqnarray}
At $\vec{k}^2 = 0$, the sum, as noted before, becomes 
\begin{eqnarray}
\sum_{n \geq 1} \int dz h(z) \langle \psi_{2n-1}(z) \psi_{2n-1}(Z) \rangle = \sum_{n \geq 1} \kappa \int dz h(z) \frac{1}{\kappa h(z)} \delta(z-Z) = 1.
\end{eqnarray}
There is conflicting evidence as to whether, and to what extent, the resulting degeneracy is a flaw. Lattice results for the negative parity octets \cite{LeeAlexandru}, suggest a different behavior. However, both chiral perturbation theory \cite{Narodetskii}, and field theoretical quark models \cite{ChiangYang}, suggest that the deviation from degeneracy is not so severe. The results from these various models are reproduced in Table~\ref{tab8}, adapted from a talk by F. Lee and A. Alexandru at Lattice 2010~\cite{LeeAlexandru}.

\section{Conclusions}

We have analyzed the holographic model for the three-flavor baryons using the newly proposed Chern-Simons term~\cite{Chern} 
in the presence of symmetry breaking effects in bulk.  The new Chern-Simons term obeys all the strictures required by the chiral anomaly, and generates the key hypercharge constraint in the collective quantization of the baryon spectra.  For the three-flavor under consideration, the vector and axial-vector currents are also found to obey strict vector dominance as originally noted for the
two-flavor case. The results for the many of the bulk parameters of the octet and decuplet baryons are reproduced with marked quantitative improvement with respect to the two-flavor case. We have also analyzed some bulk properties of the excited octet
baryons with comparison to some existing models and lattice simulations, which maybe accessible in future experiments.

\clearpage
\begin{table}
\caption{Values of magnetic moment   $\mu$   for the octet (in units of the Bohr nuclear magneton,   $\mu_N$  )} \label{tab1}
\begin{ruledtabular}
\begin{tabular}{|c||c|c|c|c|c|}
& Q & Y &  SU(3)  Symmetry & Broken Symmetry & Empirical Values\cite{PDG} \\ \hline
 N  &  0  &  1  &  -1.6667  & -1.6292 &  -1.91  \\
 P  &  1  &  1  &  2.2224  &  2.2619  &  2.79287  \\ 
  $\Sigma^{+}$   &  1  &  0  &  2.2224  & 2.2595 &  2.458  $\pm$  0.010  \\ 
  $\Sigma^{-}$   &  -1  &  0  &  -0.5556  & -0.5202 &  -1.16  \\ 
	$\Sigma^{0}$ &  0 & 0 & 0.6211 & 0.6494 & \\
	$\Lambda^{0}$ & 0 & 0 & 0.6211 & 0.6454 & \\
  $\Xi^0$   &  1  &  -1  &  -1.6668  & -1.6516 &  -1.25  \\ 
  $\Xi^{-}$   &  -1  &  -1  &  -0.5556  & -0.5405 &  -0.69  \\ 
\end{tabular}
\end{ruledtabular}
\end{table}

\begin{table}
\caption{Values of magnetic moment  $\mu$  for the decuplet in units of $\mu_B$} \label{tab2}
\begin{ruledtabular}
\begin{tabular}{|c||c|c|c|c|c|c|}
&  Q  &  Y  &  SU(3)  Symmetry & Broken Symmetry & Measurement & Lattice Predictions \\ \hline
 $\Delta^{++}$  &  2  &  1  &  4.7406  & 4.8189 & N/A & 4.52 $\pm$ 0.51 $\pm$ 0.45\cite{Bosshard}, 4.91(61)\cite{Leinweber} \\ 
 $\Delta^{+}$  &  1  &  1  &  2.37028  & 2.4459 & N/A & 2.7 $\pm$ 1.5 \cite{Kotulla}, 2.46(31)\cite{Leinweber}\\ 
 $\Delta^{0}$  &  0  &  1  &  0  & 0.0729 & N/A & 0.00 \cite{Leinweber} \\ 
 $\Delta^{-}$  &  -1  &  1  &  -2.37028  & -2.3001 & N/A & -2.46(31)\cite{Leinweber} \\ 
 $\Sigma^{*+}$  &  1  &  0  &  2.37028  & 2.4382 & N/A & 2.55(26)\cite{Leinweber} \\ 
 $\Sigma^{*0}$  &  0  &  0  &  0  & 0.0233 & N/A & 0.27(5)\cite{Leinweber} \\ 
 $\Sigma^{*-}$  &  -1  &  0  &   -2.37028  & -2.3095 & N/A & -2.02(18)\cite{Leinweber} \\ 
 $\Xi^{*0}$  &  0  &  -1  &  0  & 0.0567  & N/A & 0.46(7)\cite{Leinweber} \\
 $\Xi^{*-}$  &  -1  &  -1  &  -2.37028  & -2.3185 & N/A & -1.68(12)\cite{Leinweber} \\
 $\Omega^{-}$  &  -1  &  -2  &  -2.37028  & -2.2935 & -2.02 $\pm$ 0.05\cite{PDG} & -1.40(10)\cite{Leinweber} \\ 
\end{tabular}
\end{ruledtabular}
\end{table}

\clearpage
\begin{table}
\caption{Axial transition constants for the octet with and without symmetry breaking, compared to measured values} \label{tab3}
\begin{ruledtabular}
\begin{tabular}{|c||c|c|c|}
& SU(3) Symmetry & Broken Symmetry & Empirical Values \cite{Borasoy}
\\ \hline
 N $\rightarrow$ P  &  1.1578  & 1.1484 &    1.26  \\ 
 $\Sigma^{-} \rightarrow \Lambda$  &  0.6023 & 0.6031  &  0.61 $\pm$ 0.02  \\
 $\Xi^{-} \rightarrow \Xi^0$  &  0.3279  &  0.3279 &   \\
 $P \rightarrow \Lambda$  &  -0.8125  &  -0.802  &  -0.92  \\ 
 $\Sigma^{-} \rightarrow$ N  &  0.328  & 0.3284   &  0.39  \\ 
\end{tabular}
\end{ruledtabular}
\end{table}

\begin{table}
\caption{Axial transition constants for the decuplet with and without symmetry breaking} \label{tab4}
\begin{ruledtabular}
\begin{tabular}{|c||c|c|c|}
& SU(3) Symmetry & Broken Symmetry & PCQM\cite{Axial} \\ \hline
 $\Delta^{-} \rightarrow \Delta^0$  &  1.2117  &1.2132 & 1.52113 \\

 $\Delta^0 \rightarrow \Delta^{+}$  &  1.3991  & 1.4008  & 1.75645 \\

 $\Sigma^{* -} \rightarrow \Delta^0$  &  0.6996  & 0.6989 & 0.87823 \\

 $\Sigma^{* 0} \rightarrow \Delta^{+}$  &  0.9892  & 0.9884 & 1.242 \\ 

 $\Sigma^{* +} \rightarrow \Delta^{++}$  &  1.2117  & 1.2111 & 1.52113 \\

 $\Sigma^{* -} \rightarrow \Sigma^{* 0}$  &   0.9892  & 0.991 & 1.242 \\ 

 $\Xi^{*0} \rightarrow \Sigma^{*+}$  &  1.3991  & 1.3982 & 1.75645 \\ 

 $\Omega^{-} \rightarrow \Xi^{* 0}$  &   1.2117  & 1.2111 & 1.52113 \\ \hline
\end{tabular}
\end{ruledtabular}
\end{table}

\begin{table}
\caption{Axial couplings for decuplet to octet transitions} \label{tab5}
\begin{ruledtabular}
\begin{tabular}{|c||c|}
&  SU(3)  Symmetry \\ \hline
 $\Delta^0 \rightarrow$ P  &  -0.9599  \\
 $\Delta^{-} \rightarrow$ N  &  -1.6626  \\ 
 $\Sigma^{*0} \rightarrow \Sigma^{+}$  &  0.6788  \\
 $\Sigma^{*-} \rightarrow \Lambda$  &  -1.1757  \\
 $\Xi^{*-} \rightarrow \Xi^{0}$  &  0.9599 \\
\end{tabular}
\end{ruledtabular}
\end{table}

\clearpage
\begin{table}
\caption{Electric charge radii of decuplet baryons (fm)} \label{tab6}
\begin{ruledtabular}
\begin{tabular}{|c||c|c|c|}
& prediction & field theoretical quark model \cite{Sahoo} & $1/N_c$ analysis \cite{Lebed} \\ \hline
$\langle r^2 \rangle_{\Delta^{++}}^{1/2}$ & 1.109 & 1.086 & 1.005\\
$\langle r^2 \rangle_{\Delta^{+}}^{1/2}$ & .7844 & .9055 & 1.005 \\
$\langle r^2 \rangle_{\Delta^{0}}^{1/2}$ & 0 & .4 & 0 \\
$\langle r^2 \rangle_{\Delta^{-}}^{1/2}$ & .7844 & .9165 & 1.005 \\
$\langle r^2 \rangle_{\Sigma^{*+}}^{1/2}$ & .7844 & .9849 & 1.005 \\
$\langle r^2 \rangle_{\Sigma^{*-}}^{1/2}$ & .7844 & .9165 & .9194 \\
$\langle r^2 \rangle_{\Sigma^{*0}}^{1/2}$ & 0 & .5831 & .3563 \\
$\langle r^2 \rangle_{\Xi^{*0}}^{1/2}$ & 0 & .7 & .494 \\
$\langle r^2 \rangle_{\Xi^{*-}}^{1/2}$ & .7844 & .9055 & .8319 \\
$\langle r^2 \rangle_{\Omega^{-}}^{1/2}$ & .7844 & .8832 & .7436
\end{tabular}
\end{ruledtabular}
\end{table}

\clearpage
\begin{table}
\caption{Axial charges of excited nucleon states} \label{tab7}
\begin{ruledtabular}
\begin{tabular}{|c||c|c|c|}
($n_{\rho}$, $n_z$) & possible state identification \cite{Baryon} & $g_A$ with SU(3) Symmetry  & RCQM \cite{Excited} \cite{Excitedone}  \\ \hline
(1,0) & $N(1440)$ & 1.482 & 1.16 \\
(0,1) & $N^{*}(1535)$ & 0.595 & 0.02 (EGBE), 0.13 (OGE) \\
(1,1) & $N^{*}(1655)$ & 0.769 & 0.51 (EGBE), 0.44 (OGE) \\
(0,2) & $N(1710)$ & 0.5204 & 0.35 \\
(2,1) & $N^*(2090)$ & 0.942 &  \\
(0,1) & $\Xi(1690)$ & -0.423 & -0.23 \\
(1,0) & $\Sigma(1660)$ & 0.529 & 0.69 \\
\end{tabular}
\end{ruledtabular}
\end{table}
\clearpage

\begin{table}
\caption{Predictions for Magnetic Moments of Negative Parity Octet Baryon Excitations, Compared to Octet Baryons} \label{tab8}
\begin{ruledtabular}
\begin{tabular}{|c|c|c|c|c|c|}
State & Holographic Value (above) & Experimental Value $\mu$ (units of $\mu_N$)  & FTQM \cite{ChiangYang} & Lattice Prediction \cite{LeeAlexandru} & $\chi$PT \cite{Narodetskii}  \\ \hline
P & 2.26 & 2.79 & & & \\
P^{*}(1/2-) & 2.26 & N/A & 1.89 & -1.0 & 1.1  \\ 
N & -1.63 & -1.91 & & & \\
N^{*}(1/2-) (1535) & -1.63 & N/A &  -1.28 & -0.5 & -0.25 \\
$\Lambda_0$ & -0.65 & -0.61 & & & \\ 
$\Lambda^{*}_0 (1/2-)$ & -0.65 & N/A &  +0.28 & -0.3 & -0.29  \\ 
$\Sigma^{0}$ & 0.65 & 0.65 & & & \\
$\Sigma^0(1/2-)$ & 0.65 & N/A & -0.5 & N/A & N/A \\ 
\end{tabular}
\end{ruledtabular}
\end{table}

\section{Acknowledgements}
This work is supported by the U.S. Department of Energy under Contract No.
DE-FG-88ER40388. PHCL is supported by the Croucher Foundation postdoctoral fellowship.

%\end{document}

\end{document}